\documentclass[12pt]{article}

\usepackage{amssymb}
\usepackage{amsmath}
\usepackage{epsfig}

\begin{document}

\title{\bf Gravireggeons and transplanckian scattering in models
with one extra dimension}

\author{A.V. Kisselev\thanks{E-mail: alexandre.kisselev@mail.ihep.ru} \
and V.A. Petrov\thanks{E-mail: vladimir.petrov@mail.ihep.ru} \\
\small Institute for High Energy Physics, 142281 Protvino, Russia}

\date{}

\maketitle

\thispagestyle{empty}

\bigskip

\begin{abstract}
The inelastic scattering of the brane fields induced by
$t$-channel gravireggeons exchanges in the RS model with a small
curvature $\kappa$ is considered, and the imaginary part of the
eikonal is analytically calculated. It is demonstrated that the
results can be obtained from the corresponding formulae previously
derived in the ADD model with one extra dimension of the size
$R_c$ by formal replacement $R_c \rightarrow (\pi \kappa)^{-1}$.
The inelastic cross section for the scattering of ultra-high
neutrino off the nucleon is numerically estimated for the case
$\kappa \ll \bar{M}_5 \sim 1$ TeV, where $\bar{M}_5$ is a reduced
Planck scale in five warped dimensions.
\end{abstract}

%%%%%%%%%%%%%
% Main text %
%%%%%%%%%%%%%

\section{Introduction}

In our previous papers~\cite{Kisselev:04,Kisselev:04*}, we
calculated the contribution of Kaluza-Klein (KK) gravireggeons
into the scattering of four-dimensional SM particles in a model
with $d$ compact extra spatial dimensions (the ADD
model~\cite{Arkani-Hamed:98}). The results were applied to the
scattering of cosmic neutrinos off nucleons at superplanckian
neutrino energy $E_{\nu}$.

At $10^{8} \mathrm{\ GeV} < E_{\nu} < 10^{12} \mathrm{\ GeV}$, the
cross sections related with gravity interactions appeared to be
compatible with (larger than) SM cross sections at $d \leqslant 3
\div 4$, depending on $E_{\nu}$. The gravitational part of the
cross section induced by the gravireggeon exchange rises rapidly
with a decrease of $d$. For instance, for $d=2$, it is
approximately two orders of magnitude larger than the SM
contribution to the cross section~\cite{Kisselev:04}, if the
gravity scale is chosen to be $1\div2$ TeV.

Unfortunately, present astrophysical bounds~\cite{Hannestad:03}
rule out the possibility $d=2$ and significantly restrict the
parameter space for $d=3$. The case $d=1$ is completely excluded
since a radius of a single extra dimension, $R_c$, exceeds the
size of the Universe, if we insist that a fundamental gravity
scale in five dimensions, $\bar{M}_5$, should be 1 TeV or so. It
follows from the relation $R_c^d \sim
\bar{M}_{Pl}^2/\bar{M}_D^{2+d}$, with $\bar{M}_D$ being a
$D$-dimensional reduced Planck scale ($D=4+d$).

However, the above mentioned astrophysical bounds do not apply to
the extra dimensions with a warped metric. In the present paper we
consider a model of gravity in a slice of a 5-dimensional Anti-de
Sitter space (AdS$_5$) with a single extra dimension compactified
to the orbifold $S^1/Z_2$ (the RS
model~\cite{Randall:99,Randall:99*}). We consider a special case
when a curvature of the metric $\kappa$%
\footnote{The Ricci curvature invariant for this AdS$_5$ space is
given by $\mathcal{R}^{(5)} = -20 \, \kappa^2$.}
is much smaller than the gravity scale $\bar{M}_5$.

We demonstrate that in such a limit the expression for inelastic
cross section for a collision of the brane particles in
\emph{warped five dimensions} can be obtained from the analogous
expression previously derived in \emph{five flat dimensions} by a
formal substitution $R_c \rightarrow 1/(\pi \kappa)$. Then
numerical calculations  show that the gravity (gravireggeon)
contribution to the scattering of the brane fields should dominate
the SM contribution even for rather large $\bar{M}_5$.

\section{RS model with a small curvature}
\label{sec:RS}

In the RS model,  the warped metric is of the form:
\begin{equation}\label{02}
ds^2 = e^{-2 \kappa |y|} \, \eta_{\mu \nu} \, dx^{\mu} \, dx^{\nu}
+ dy^2.
\end{equation}
Here $y = r \theta$ ($-\pi \leqslant \theta \leqslant \pi$), $r$
being a "radius" of extra dimension, and parameter $\kappa$
defines the scalar  curvature of the space.

We are interested in so-called RS1 model~\cite{Randall:99} which
has two 3-dimensional branes with equal and opposite tensions
located at the point $y = \pi r$ (called the \emph{TeV brane}, or
\emph{visible brane}) and at $y = 0$ (referred to as the
\emph{Plank brane}). If $k > 0$, then the tension on the TeV brane
is negative, whereas the tension on the Planck brane is positive.
All SM fields are constrained to the TeV 4D brane, while the
gravity propagates in all five dimensions (bulk).

From an effective 4-dimensional action one can derive the
relation~\cite{Randall:99}:
\begin{equation}\label{04}
\bar{M}_{Pl}^2 = \frac{\bar{M}_5^3}{\kappa} \left( 1 - e^{-2 \pi
\kappa r} \right),
\end{equation}
which means that $\kappa \sim \bar{M}_5 \sim \bar{M}_{Pl}$ in this
case, $\bar{M}_5$ being a 5-dimensional reduced Planck scale.

The kinetic energy in the visible brane action is not canonically
normalized. After re-scaling the fields, the warp factor appears
in a mass term:
\begin{equation}\label{06}
m \rightarrow e^{-\pi \kappa r} m.
\end{equation}
As a result, the masses of the Kaluza-Klein (KK) graviton
excitations are given by
\begin{equation}\label{08}
m_n = x_n \, |\kappa| \, e^{-\pi \kappa r}, \qquad n=1,2 \ldots,
\end{equation}
where $x_n$ are zeros of the Bessel function $J_1(x)$. Note, here
and in what follows we are interested in a case $\kappa r > 1$,
neglecting terms $\sim e^{-\pi \kappa r}$ with respect to 1.

The interaction Lagrangian on the brane with a negative tension
looks like%
\footnote{We do not consider radion field here because for
high-energy $t$-channel exchanges they are irrelevant.}
\begin{equation}\label{10}
\mathcal{L} = - \frac{1}{\bar{M}_{Pl}} \, T^{\mu \nu} \,
h^{(0)}_{\mu \nu} - \frac{1}{\Lambda_{\pi}} \, T^{\mu \nu} \,
\sum_{n=1}^{\infty} h^{(n)}_{\mu \nu}.
\end{equation}
Here $T^{\mu \nu}$ is the energy-momentum tensor of the matter on
the brane, $h^{(n)}_{\mu \nu}$ is the graviton field with a
KK-number $n$, and
\begin{equation}\label{12}
\Lambda_{\pi} = \bar{M}_{Pl} \, e^{-\pi \kappa r_c} \simeq \left(
\frac{\bar{M}_5}{\kappa} \right)^{3/2} \, \frac{m_n}{x_n}
\end{equation}
is the physical scale on the TeV brane. It can be chosen as small
as 1 TeV  for a thick slice of the AdS$_5$,
\begin{equation}\label{14}
r \simeq 12/\kappa \simeq 60 \, l_{Pl}.
\end{equation}
We see from \eqref{10} that couplings of all massive states are
suppressed by $\Lambda_{\pi}^{-1}$ only, while the zero mode
couples with usual strength defined by the reduced Planck mass
$\bar{M}_{Pl} = M_{Pl}/\sqrt{8 \pi}$.

The relation \eqref{14} guarantees that the masses of lowest
graviton KK excitations~\eqref{08} are closed to one TeV. Thus,
the phenomenology of models with non-factorizable metric is
associated with the resonant KK spectrum in the TeV
region~\cite{Davoudiasl:00}.

Let us note, that because of the warp factor $e^{-2 \kappa |y|}$
on the TeV brane, the coordinates $x^{\mu}$ are not Galilean. One
can, however, introduces the Galilean coordinates $z^{\mu} =
x^{\mu} e^{- \pi \kappa r}$ and rewrite both the gravitational
field and the energy-momentum tensor in these coordinates (see,
for instance, reviews \cite{Rubakov:01}). Then the warp factor is
equal to 1 at the negative tension brane, and a correct
determination of the masses on this brane can be
achieved~\cite{Boos:02}. By calculating the zero mode sector of
the effective theory one thus obtains:
\begin{equation}\label{16}
\bar{M}_{Pl}^2 = \frac{\bar{M}_5^3}{\kappa} \left( e^{2 \pi \kappa
r} - 1 \right).
\end{equation}

In such a case, we have the following mass spectrum on the
negative tension brane:
\begin{equation}\label{18}
m_n = x_n \, |\kappa|, \qquad n=1,2 \ldots.
\end{equation}
To get $m_n \sim 1$ TeV, the parameters of the model are usually
taken to be $\kappa \sim \bar{M}_5 \sim 1$ TeV. Because of the
relation
\begin{equation}\label{20}
x_n = \pi \, \Big( n + \frac{1}{4} \Big) + \frac{3}{8} \Big[ \pi
\, \Big( n + \frac{1}{4} \Big) \Big]^{-1} + \mathrm{O \left(
n^{-2} \right)},
\end{equation}
the KK states are equally spaced at large $n$.%
\footnote{The first four values of $x_n$ are 3.83, 7.02, 10.17,
and 13.32.}
The interaction Lagrangian is defined by Eq.~\eqref{10} with the
expression of $\Lambda_{\pi}$ to be%
\footnote{The KK gravitons have a universal coupling at both small
and large $\kappa r$~\cite{Giudice:04}. Remember, we do not
consider the intermediate region $\kappa r \sim 1$.}
\begin{equation}\label{22}
\Lambda_{\pi} \simeq \frac{\bar{M}_5^{3/2}}{\kappa^{1/2}} = \left(
\frac{\bar{M}_5}{\kappa} \right)^{3/2} \, \frac{m_n}{x_n}.
\end{equation}

The mass scales of the parameters of the RS model are quite
different (Planck scale in the first case and TeV scale in the
second case), but a particle phenomenology is similar. Indeed, the
substitution $\kappa \rightarrow \kappa e^{-\pi \kappa r}$,
$\bar{M}_5 \rightarrow \bar{M}_5 \,  e^{-\pi \kappa r}$ provides
us with the same mass spectrum of the massive gravitons and the
same coupling of the KK gravitons to the SM fields.
%In this case, relation \eqref{16} turns into relation \eqref{04}.

Nevertheless, the correct statement is that the masses of the KK
gravitons, as seen by an observer living on the brane with the
negative tension, are defined by Eq.~\eqref{18}~\cite{Rubakov:01}.
On the other brane, their values are defined by Eq.~\eqref{08}.
Moreover, all bulk fields, not only gravitons, look differently to
observers on different branes. The observed masses for brane
fields coincide with their Lagrangian values and do not depend on
coordinate rescaling, if covariant equations and invariant
distances are used~\cite{Grinstein:01}.

Generally speaking, we have \emph{three} dimensional parameters in
the RS model: fundamental gravity scale in 5 dimensions,
$\bar{M}_5$, the curvature scale, $\kappa$, and the size of extra
dimension, $r$. They obey only \emph{two} conditions. Indeed, to
get TeV physics, one fixes $\bar{M}_5$ to be one or few TeV. Then
we can regard Eq.~\eqref{12} as a relation between free parameters
$\kappa$ and $r$ at fixed value of $\bar{M}_5$.%
\footnote{Some values of $\bar{M}_5$ and $\kappa$  which result in
an unnaturally large coupling constant $\Lambda_{\pi}$ should be
avoided in order not to introduce a new mass scale in the theory.}

Thus, there is a possibility to consider a case in which $\kappa$
is larger than $r^{-1}$, but is much smaller than $\bar{M}_5$. It
was recently demonstrated in  Ref.~\cite{Giudice:04}, where the
warp factor in the line element was chosen to be $e^{2\kappa |y|}$
instead of $e^{-2\kappa |y|}$:
\begin{equation}\label{24}
ds^2 = e^{2 \kappa |y|} \, \eta_{\mu \nu} \, dx^{\mu} \, dx^{\nu}
+ dy^2.
\end{equation}
The brane located at the point $y=0$ has now the negative tension.%
\footnote{This choice of the warp factor is equivalent to a
replacement $\kappa \rightarrow -\kappa$ in \eqref{02}, and the
branes are interchanged. Note that Eq.~\eqref{04} turns into
Eq.~\eqref{16} under such a replacement.}
This brane is regarded as the visible brane, while the Planck
brane is located at $y=\pi r$. The coordinates $x^{\mu}$ are
Galilean on the visible brane. It is not surprising that the
relations \eqref{16} and \eqref{18} are reproduced in this scheme.

Following \cite{Giudice:04}, the mass splitting ($\simeq \pi
\kappa $) can be chosen to be smaller than the energy resolution
of collider experiments. We take $\pi \kappa = 50$ MeV for
phenomenological purposes. Then $\kappa r \simeq 9.7$, that
corresponds to $r \simeq 0.61$ MeV$^{-1}$ $\simeq 120$ fm, and the
mass of the lightest KK excitation is $m_1 = 60.5$ MeV. The
coupling constant, $\Lambda_{\pi} \simeq (\bar{M}_5/1
\mathrm{TeV})^{3/2} \, 141$ TeV, is two orders of magnitude larger
than in the usually adopted phenomenological
scheme~\cite{Davoudiasl:00}.

There are restrictions on the parameters of the RS model. In
Refs.~\cite{Davoudiasl:00}, an upper and lower bound on the ratio
$\kappa/\bar{M}_{Pl}$ was obtained based on Eqs.~\eqref{04} and
\eqref{08} (assuming that $\kappa \sim \bar{M}_5 \sim
\bar{M}_{Pl}$). We will derive analogous bounds on the ratio
$\kappa/\bar{M}_5$, when the SM fields are on the negative tension
brane with the Galilean coordinates, and, consequently, relations
\eqref{16}, \eqref{18} are valid (assuming that $\kappa \lesssim
\bar{M}_5 \sim 1$ TeV).

We exploit the ideas used in the above mentioned paper
\cite{Davoudiasl:00}. The solution for the metric \eqref{02} can
be trusted if the 5-dimensional scalar curvature,
$\mathcal{R}^{(5)} = -20 \, \kappa^2 e^{2 \pi \kappa r}$, obeys
the inequality $|\mathcal{R}^{(5)}| < \bar{M}_5^2 \, e^{2 \pi
\kappa r}$, that results in the condition  $\kappa/\bar{M}_5
\lesssim 0.2$.

The D3-brane tension $\tau$ in the heterotic string theory is
given by~\cite{Polchinski:98}
\begin{equation}\label{26}
\tau_3 = \frac{M_s^4}{g_s (2\pi)^3},
\end{equation}
where $M_s = (\alpha')^{-1}$ is the string scale, and $g_s$ is the
string coupling constant. the low-energy action in the strongly
coupled heterotic string theory in ten dimensions looks
like~\cite{Polchinski:98}:
\begin{equation}\label{28}
S = \int d^{10}x \left[ \frac{M_s^8}{(2\pi)^7 \, g_s^2}
\mathcal{R} + \frac{1}{4} \, \frac{M_s^6}{(2\pi)^7 \, g_s} \, F^2
+ \cdots \right].
\end{equation}
After compactification of ten-dimensional action to four
dimensions  the coefficient of $\mathcal{R}$ and $(1/4) F^2$
should be  identified with $1/(16 \pi G_N)$ and $1/g_{_G}^2$,
respectively, where $g_{_G}$ is a 4-dimensional gauge coupling
taken at the string scale $M_s$. Let us first assume that all six
extra dimensions are compact ones. By performing T-duality to six
dimensions, one then obtains~\cite{Arkani-Hamed:98}:
\begin{equation}\label{30}
g_s = \frac{g_{_G}^2}{2\pi}.
\end{equation}

By compactifying the action \eqref{28} to five warp dimensions, we
get:
\begin{equation}\label{32}
\bar{M}_5^3 = \frac{2 V_5 M_s^8}{g_s (2\pi)^7},
\end{equation}
where $V_5$ is a volume of five-dimensional manifold with
non-factorizable metric. Let us now assume that the
ratio~\eqref{30} remains valid. Then, taking five extra dimensions
to have a common radius $R_c = M_s^{-1}$, we find:
\begin{equation}\label{34}
M_s = \left( \frac{g_{_G}^4}{2} \right)^{1/3} \, \bar{M}_5.
\end{equation}

On the other hand, the tension of the 3-branes in the RS model
is~\cite{Randall:99}
\begin{equation}\label{36}
|\tau| = 24 \, \bar{M}_5^3 \, \kappa.
\end{equation}
Requiring $|\tau| = \tau_3$, one gets from \eqref{26}, \eqref{34}
and \eqref{36} that $\kappa/\bar{M}_5 \simeq 6.1 \cdot 10^{-4}$
$(1.3 \cdot 10^{-5})$ for $\alpha_{_G} =g_{_G}^2/4\pi \simeq 0.1$
$(0.01)$. As one can see, the value of the ratio
$\kappa/\bar{M}_5$ depends on which of the SM gauge couplings are
chosen to represent $g_{_G}$. We take the following region for a
phenomenological analysis:
\begin{equation}\label{38}
10^{-5} \leqslant \frac{\kappa}{\bar{M}_5} \leqslant 0.1.
\end{equation}
For $\bar{M}_5 = 1$ TeV, Eq.~\eqref{38} corresponds to $10
\mathrm{\ MeV} \leqslant \kappa \leqslant 0.1 \mathrm{\ TeV}$.
Remember that the fundamental mass scale is related with the
Planck mass by Eq.~\eqref{16}, while the masses of the KK
excitations are given by Eq.~\eqref{18}. In what follows, we will
be interested in a case when the ration $\kappa/\bar{M}_5 $ is
closed to the lower end of the range \eqref{38}, and $\kappa \ll
\bar{M}_5 \sim 1$ TeV.

Note, the RS model with the small curvature may be regarded as a
small distortion of the compactified flat space with one large
extra dimension. Such space warping gives a model which has the
ultraviolet properties of the ADD model with a single extra
dimensions~\cite{Giudice:04}, but it evades the contradiction with
available astrophysical bounds.

\section{Eikonal in flat and warp five dimensions}
\label{sec:eikonal}

Now let us consider a scattering of two \emph{point-like} brane
particles (say, lepton-quark or quark-quark scattering) in the
transplanckian kinematical region
\begin{equation}\label{50}
\sqrt{s} \gg M_D, \qquad s \gg -t,
\end{equation}
$t = -q_{\bot}^2$ being four-dimensional momentum transfer. More
realistic case of neutrino-proton interactions will be studied in
the next Section.

In the eikonal approximation an elastic scattering amplitude in
the kinematical region \eqref{50} is given by the sum of reggeized
gravitons in $t$-channel. So, we assume that both massless
graviton and its KK massive excitations lie on linear Regge
trajectories. Due to a presence of the extra dimension, we come to
splitting of the Regge trajectory \eqref{28} into a leading vacuum
trajectory
\begin{equation}\label{52}
\alpha_0(t) \equiv \alpha_{grav}(t) = 2 + \alpha_g' t,
\end{equation}
and an infinite sequence of secondary, ``KK-charged'',
gravireggeons~\cite{Petrov:02,Kisselev:04}:
\begin{equation}\label{54}
\alpha_n(t) = 2 + \alpha_g' t  - \alpha_g' \,  m_n^2, \quad n
\geqslant 1.
\end{equation}
The string theory implies that the slope of the gravireggeon
trajectory is universal for all $s$, and $\alpha_g' = \alpha' =
1/M_s^2$.

Let us first consider the scattering of the brane fields in a
model with $d$ \emph{flat} compact extra
dimensions~\cite{Arkani-Hamed:98}. In the ADD model the masses of
the KK gravitons are given by $m_n^2=n^2/R_c^2$, where $n^2 =
n_1^2 + \cdots + n_d^2$, and $R_c$ is the compactification radius
of the extra dimensions. The coupling of both zero and massive
modes to colliding particles are suppressed by the Planck scale.
Therefore, the Born amplitude looks like
\begin{equation}\label{56}
A^B_{_{ADD}}(s,t) = \frac{\pi \, \alpha_g' \, s^2}{2
\bar{M}_{Pl}^2} \sum_{n_1, \ldots n_d} \Big[ i - \cot
\frac{\pi}{2} \alpha_n(t) \Big] \left( \frac{s}{s_0}
\right)^{\alpha_n(t) - 2}.
\end{equation}
It defines $\chi(s,b)$, the eikonal in the impact space.

Let us consider the imaginary part of the eikonal in which the
zero mode contribution is negligible. The analytical expression
for $\text{Im}\,\chi(s,b)$ was derived in~\cite{Kisselev:04}:
\begin{eqnarray}\label{60}
\text{Im}\,\chi_{_{ADD}}(s,b) &=& \frac{s \, \alpha_g'}{16
\bar{M}_{Pl}^2 \, R_g^2 (s)}
\exp \Big[ -b^2/ 4 R_g^2 (s) \Big] \nonumber \\
&\times& \left\{ 1 + 2 \sum_{n=1}^{\infty} \, \exp \Big[ -n^2 \,
\alpha_g' \, \ln (s/s_0)/R_c^2 \Big] \right\}^d,
\end{eqnarray}
where
\begin{equation} \label{62}
R_g(s) = \sqrt{\alpha_g' \, [\ln (s/s_0) + b_0]}
\end{equation}
is a gravitational slope. Since $b_0 = \mathrm{O}(1)$, it can be
neglected at large $s$~\cite{Kisselev:04}.

The sum in Eq.~\eqref{60} is one of the Jacobi
$\theta$-functions~\cite{Erdelyi:III}:
\begin{equation}\label{64}
\theta_3 (0, p) = 1 + 2 \sum_{k=1}^{\infty} p^{k^2},
\end{equation}
with
\begin{equation}\label{66}
p = \exp \Big[ -\alpha_g' \ln (s/s_0)/R_c^2 \Big].
\end{equation}
By using unimodular transformation of the
$\theta_3$-function~\cite{Erdelyi:III} (known also as Jacobi
imaginary transformation) one can obtain the following asymptotic
of $\theta_3 (0, p)$ for large extra dimensions:
\begin{equation}\label{68}
\theta_3 (0, p) \Big|_{R_c^2 \gg \alpha_g' \ln (s/s_0)} \simeq
\sqrt{\frac{\pi \, R_c^2}{\alpha_g' \ln (s/s_0)}}.
\end{equation}

As a result, we get that in a flat metric with $d$ extra compact
dimensions the imaginary part of the eikonal is of the
form~\cite{Kisselev:04}:
\begin{equation}\label{70}
\text{Im} \, \chi_{_{ADD}}(s,b) \simeq \frac{1}{16 \pi^{d/2} \big[
\ln (s/s_0) \big]^{(2+d)/2}} \, \frac{s}{\bar{M}_D^2} \, \left(
\frac{M_s}{2 \bar{M}_D} \right)^d \, \exp \Big[ -b^2/ 4 R_g^2
(s)\Big].
\end{equation}

Now let us return to the \emph{non-factorizable} metric
\eqref{02}. According to \eqref{10}, the Born amplitude is of the
form
\begin{eqnarray}\label{86}
A^B_{_{RS}}(s,t) &=& \frac{\pi \, \alpha_g' \,
s^2}{2\bar{M}_{Pl}^2} \, \Big[ i - \cot \frac{\pi}{2} \,
\alpha_0(t)
\Big] \left( \frac{s}{s_0} \right)^{\alpha_0(t) - 2} \nonumber \\
&+& \frac{\pi \, \alpha_g' \, s^2}{2\Lambda_{\pi}^2} \sum_{n\neq0}
\Big[ i - \cot \frac{\pi}{2} \, \alpha_n(t) \Big] \left(
\frac{s}{s_0} \right)^{\alpha_n(t) - 2}.
\end{eqnarray}
The index $n$ runs over all negative and positive integers.

Zero mode contribution to the imaginary part of the eikonal (the
first term in Eq.~\eqref{86}) is negligible and can be omitted.
Then the total contribution of the massive KK excitations follows
from \eqref{86}:
\begin{equation}\label{102}
\text{Im}\,\chi_{_{RS}}(s,b) = \frac{s \, \alpha_g'}{16
\Lambda_{\pi}^2 \, R_g^2 (s)} \exp \Big[ -b^2/ 4 R_g^2 (s)\Big] \,
\sum_{n \neq 0} \, \exp [-\alpha_g' m_n^2 \ln (s/s_0)].
\end{equation}

Let us remember that we are interested in small $\kappa \ll 1$
TeV. In such a case, the sum in \eqref{102} is defined
mainly by large $n$,%
\footnote{The sum in Eq.~\eqref{102} is effectively cut, and $n
\lesssim n_{max} = (M_s/\pi \kappa) \, (\ln (s/s_0))^{-1/2} \simeq
2 \cdot 10^4 \, (\ln (s/s_0))^{-1/2}$ in our case.}
and one can put $m_n = (n + 1/2) \pi \kappa$ (see Eq.~\eqref{20}).
Then we can write
\begin{eqnarray}\label{108}
&& \sum_{n \neq 0} \, \exp [-\alpha_g' m_n^2 \ln (s/s_0)] \simeq
\sum_{n=-\infty}^{\infty} \, \exp [-(n + 1/2)^2 (\pi
\kappa)^2\alpha_g' \ln (s/s_0)] \nonumber \\
&& = \sum_{n=-\infty}^{\infty} q^{(n+1/2)^2} \equiv \theta_2 (0,
q) = \left( -\frac{\ln q}{\pi} \right)^{1/2} \, \theta_4 \left(0,
\, e^{\pi^2/\ln q} \right).
\end{eqnarray}
Here $\theta_2 (0, q)$ and
\begin{equation}\label{110}
\theta_4 (0, v) = 1 + 2 \sum_{k=1}^{\infty} (-1)^k \, v^{k^2}
\end{equation}
are the Jacobi $\theta$-functions,%
\footnote{In obtaining last term in \eqref{108}, unimodular
transformation of the $\theta_2$-function was used.}
and variable
\begin{equation}\label{112}
q = \exp \Big[ - (\pi \kappa)^2 \alpha_g' \ln (s/s_0) \Big]
\end{equation}
is introduced in \eqref{108}.

As a result, we obtain the following analytical expression for the
imaginary part of the eikonal:
\begin{eqnarray}\label{114}
&& \text{Im}\,\chi_{_{RS}}(s,b) = \frac{s \, \alpha_g'}{16
\Lambda_{\pi}^2
\, R_g^2 (s)} \, \exp \Big[ -b^2/ 4 R_g^2 (s) \Big] \nonumber \\
&& \times \left[ \pi \kappa^2 R_g^2 (s) \right]^{-1/2} \Big[ 1 + 2
\sum_{n=1}^{\infty} \, (-1)^n \exp \big( -n^2/ \kappa^2 R_g^2 (s)
\big) \Big].
\end{eqnarray}
If $\kappa \, R_g^2 (s) \ll 1$,%
\footnote{Since $\ln (s/s_0)$ rise slowly in $s$, this conditions
is satisfied if $\kappa \ll (\alpha_g')^{-1} = M_s$.}
then all terms in the sum are exponentially suppressed with
respect to unity. As for the leading term in Eq.~\eqref{114}, it
can be obtained from the expression for the imaginary part of the
eikonal in the ADD model with a single extra dimension by using
the following replacements in the KK sector:
\begin{equation}\label{116}
\bar{M}_{Pl} \rightarrow \Lambda_{\pi}, \qquad R_c \rightarrow
\frac{1}{\pi \kappa}.
\end{equation}
Indeed, for $\kappa \ll M_s$, we get from \eqref{114}:
\begin{equation}\label{118}
\text{Im}\,\chi_{_{RS}}(s,b) \Big|_{\kappa \ll M_s} \simeq
\frac{1}{16 \pi^{1/2} \big[ \ln (s/s_0) \big]^{3/2}} \, \frac{s \,
M_s}{\bar{M}_5^3}  \, \exp \Big[ -b^2/ 4 R_g^2 (s)\Big],
\end{equation}
and $\text{Im}\,\chi_{_{RS}}(s,b)$~\eqref{118} coincides with
$\text{Im}\,\chi_{_{ADD}}(s,b)$~\eqref{70} for $d=1$ up to a numerical
factor $1/2$, if we identify 5-dimensional (reduced) Planck scales
$\bar{M}_5$ in both schemes.

Note that the asymptotic of the eikonal \eqref{118} does not
depend on $\kappa$ in the limit $\kappa \ll M_s$, up to
insignificant corrections $\mathrm{O}(\exp[-M_s^2/\kappa^2 \, \ln
(s/s_0)])$. It allow us to study the dependence of the gravity
induced cross sections on the parameters $\bar{M}_5$ and $ M_s$.
In what follows, we will use $s_0 = \alpha'_g$, a scale motivated
by the string theory.

\section{Ultra-high energy neutrino-nucleon scattering induced
by gravitational interactions}
\label{sec:neutrino-nucleon}

Let us now apply our results to scattering of ultra-high energy
cosmic neutrinos off atmospheric nucleons (protons, for a
certainty). In the eikonal approximation, the neutrino-proton
inelastic cross section is
\begin{equation}\label{200}
\sigma^{\nu p}_{in}(s) = \int \!\! d^2 b \left\{ 1 - \exp \big[- 2
\text{Im} \, \chi_{\nu \! p}(s,b) \big] \right\}.
\end{equation}
with the eikonal defined by
\begin{equation}\label{202}
\chi_{\nu p}(s,b) = \frac{1}{4\pi s} \int\limits_0^{\infty}
q_{\bot} d q_{\bot} \, J_0(q_{\bot} b) \, A_{\nu p}^B(s,
-q_{\bot}^2).
\end{equation}
In the RS model with the small curvature ($\kappa \ll \bar{M}_5$,
$M_s$), the Born amplitude can be easily calculated by using
formulae obtained in Section~\ref{sec:eikonal}:
\begin{equation}\label{204}
A_{\nu p}^B(s, t) = \frac{s^2}{2 \sqrt{\pi} \bar{M}_5^3 \, M_s^2}
\, \sum_i \, \int\limits_{s_0/s}^1 \!\! dx \, x^2 \, \frac{1}{ R_g
(sx)} \, \exp \left[ t \, R_g^2 (sx) \right] F_i(x,t,\mu^2),
\end{equation}
where $F_i(x,t,\mu^2)$ is a skewed ($t$-dependent) distribution of
parton $i$ ($i = q,\ \bar{q}, \ g$) inside the proton. The mass
scale in $F_i(x,t,\mu^2)$ is defined by a large scale induced by
gravitational forces, $\mu = 1/[2 R_g(s)]$. The gravitational
interaction radius, $R_g (s)$, is introduced above \eqref{62}.

Assuming the Regge behavior, we can write
\begin{equation}\label{206}
F_i(x,t,\mu^2) = f_i(x,\mu^2) \exp \left[ t (r^2 +
\alpha'_{\mathrm{P}} \ln (1/x) \right],
\end{equation}
where $\alpha'_{\mathrm{P}}$ is the Pomeron slope, and
$f_i(x,\mu^2)$ is a standard parton distribution function (PDF) of
parton $i$ in momentum fraction $x$. We use a set of PDF's from
paper~\cite{Alekhin:02} based on an analysis of existing deep
inelastic data in the next-to-leading order QCD approximation in
the fixed-flavor-number scheme. The PDF's are available in the
region $10^{-7} < x < 1$, $2.5 \text{ GeV}^2 < Q^2 < 5.6 \cdot
10^7 \text{ GeV}^2$~\cite{Alekhin:02}.

We will use a fit from Ref.~\cite{Petrov:02*} for the radius $r$
and slope of the \emph{hard} Pomeron (remember that $\mu \sim M_s
\sim 1$ TeV):
\begin{equation}\label{208}
r^2 = 0.62 \mathrm{\ GeV}^{-2}, \qquad \alpha'_{\mathrm{P}} =
0.094 \mathrm{\ GeV}^{-2}.
\end{equation}
Since $r^2 \gg  R_g^2 (s)$ (at any conceivable $s$), a fall-off of
the eikonal in impact parameter $b$ will be mainly defined by
strong interactions (namely, by typical hadronic scale $r$ of
order 1 GeV$^{-1}$), and not by short-range gravitational forces
due to KK gravireggeons.

\begin{figure}
\centering \epsfysize=7cm \epsffile{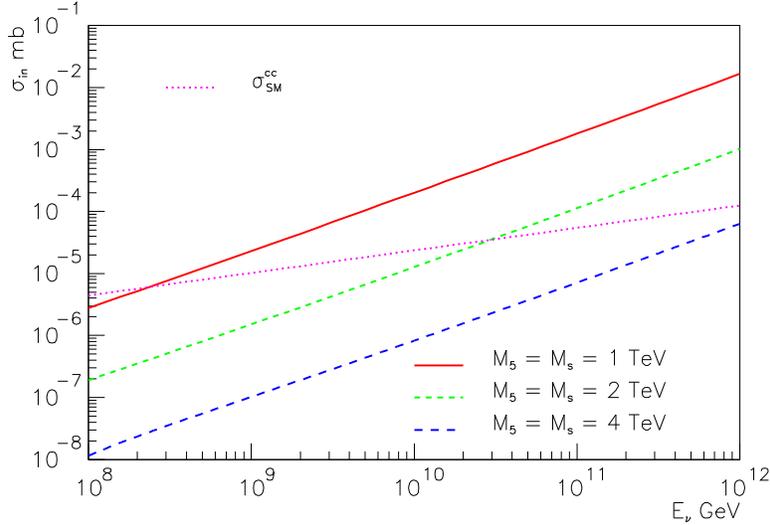}
\caption{The gravitational inelastic neutrino-proton cross-section
as a function of the neutrino energy $E_{\nu}$ for three different
values of the (reduced) fundamental scale $\bar{M}_5$ and string
scale $M_s$, which are assumed to be the same. For comparison, the
SM charged current neutrino-proton cross section is presented
(dotted curve).}
\label{fig:sigma_in_01}
\end{figure}

\begin{figure}
\centering \epsfysize=7cm \epsffile{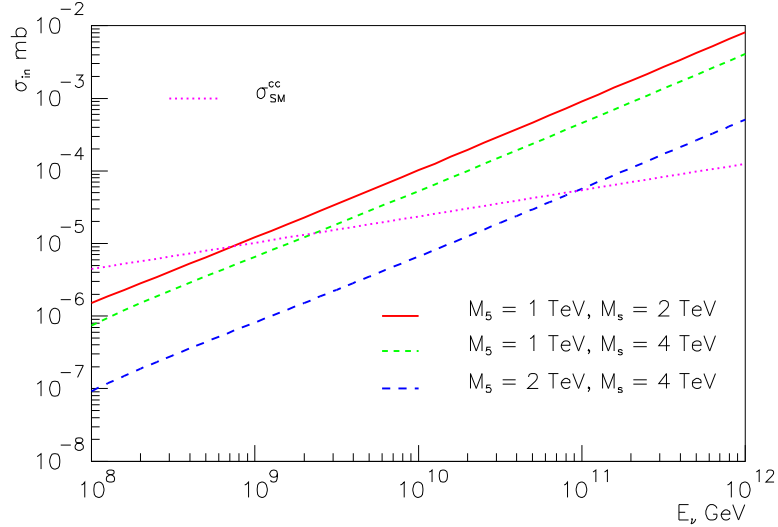}
\caption{The same as in Fig.~\ref{fig:sigma_in_01}, but the
(reduced) fundamental scale $\bar{M}_5$ is chosen to be less than
the string scale $M_s$.}
\label{fig:sigma_in_02}
\end{figure}

The inelastic cross sections induced by gravireggeons are
presented in Figs.~\ref{fig:sigma_in_01} and \ref{fig:sigma_in_02}
for different parameter sets ($\bar{M}_5, \, M_s$). In both
figures, the SM neutrino-proton charged current cross section is
also presented. An approximation for the SM cross section valid in
the range $10^{7} \, \text{GeV} \lesssim E_{\nu} \lesssim 10^{12}
\, \text{GeV}$ is taken from Ref.~\cite{Gandhi:96}.

%%%%%%%%%%%
% Figures %
%%%%%%%%%%%

\begin{figure}[h]
\centering \epsfysize=7cm \epsffile{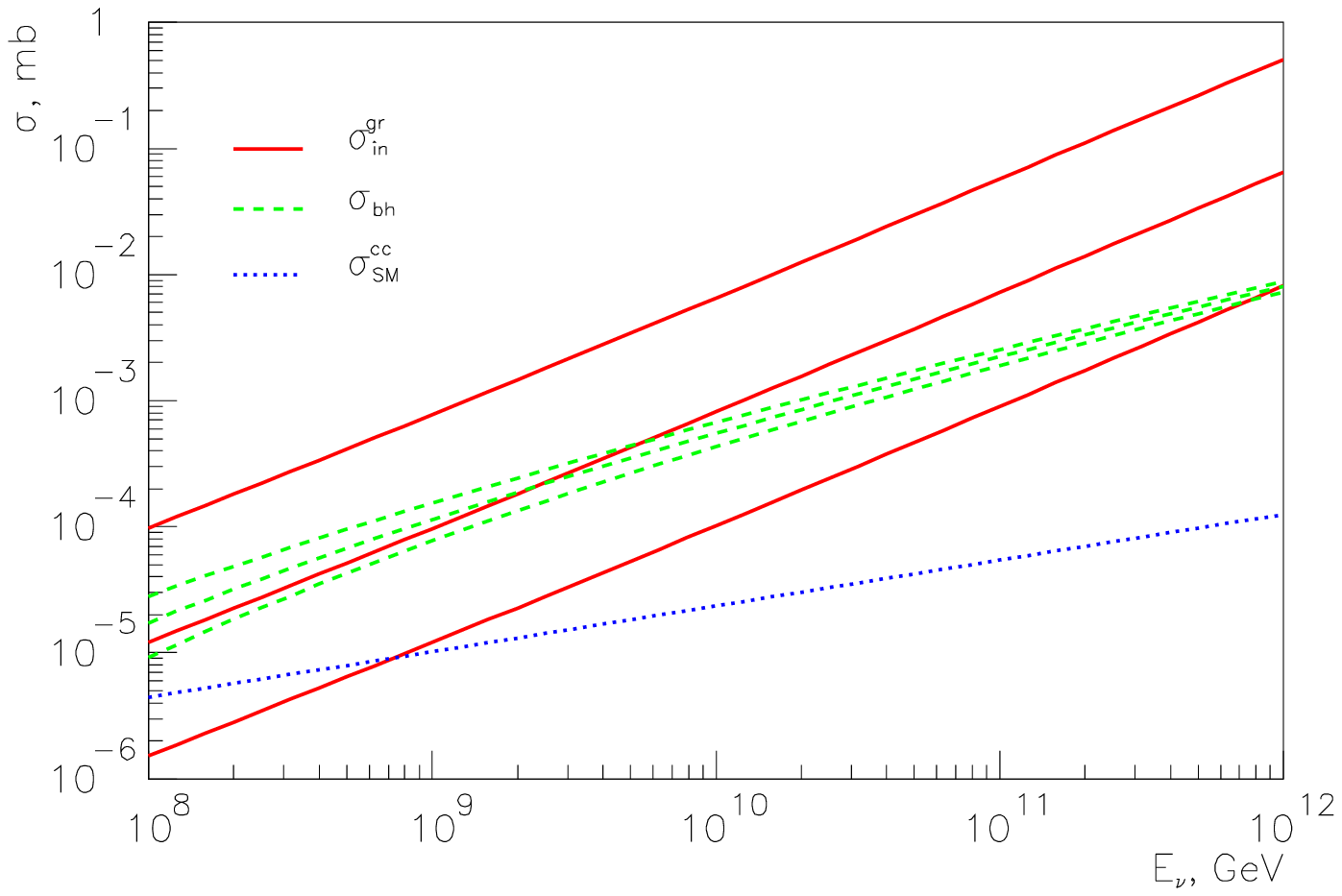}
\caption{The gravitational inelastic neutrino-proton
cross-sections (solid lines) vs. black hole production cross
sections (dash lines). The solid curves correspond to $M_s = 2$
TeV and $\bar{M}_5 = 0.25$ TeV, 0.5 TeV, 1 TeV (from the top). The
dash lines correspond to $M_5 = 0.9$ TeV (i.e. $\bar{M}_5 = 0.5$
TeV) and $M_{bh}^{min} = 0.5$ TeV, 1~TeV, 2 TeV (from the top).
The SM cross section is also shown (dotted line).}
\label{fig:sigma_in_bh}
\end{figure}

It is interesting to compare the gravitational inelastic cross
section with the black hole production cross section. Let
$\hat{\sigma}$ be the cross section of the black hole production
in the neutrino-quark (or neutrino-gluon) subprocess. Then the
black hole production cross section in the neutrino-proton
collision can be presented in the form
\begin{equation}\label{210}
\sigma_{bh}(s) = \int\limits_{(M_{bh}^{min})^2/s}^1 \! dx \,
\hat{\sigma}(\sqrt{xs}) \, \sum_i f_i(x,\tilde{\mu}^2),
\end{equation}
where $s = 2M_p E_{\nu}$ is an invariant collision energy, with
$\sqrt{xs}$ being a black hole mass $M_{bh}$. The quantity
$M_{bh}^{min}$ in \eqref{210} is a minimal value of $M_{bh}$. A
mass scale in PDF's is chosen to be $\tilde{\mu} = 1/R_S(M_{bh})$.

The cross section $\hat{\sigma}$ in Eq.~\eqref{210} is usually
taken in a simple geometrical form~\cite{Giddings:02},
\begin{equation}\label{212}
\hat{\sigma}(E) = \pi R_S^2(E),
\end{equation}
where $R_S(E)$ is the size of 5-dimensional Schwarzschild
radius~\cite{Myers:86}:
\begin{equation}\label{214}
R_S(E) = \sqrt{\frac{2}{3\pi} \, \frac{E}{M_5^3}}.
\end{equation}
As in the case of five flat dimensions, we define the fundamental
Planck scale $M_5$ to be related with the \emph{reduced} Planck
scale $\bar{M}_5$ by equation $M_5 = (2\pi)^{1/3} \, \bar{M}_5
\simeq 1.8 \, \bar{M}_5$.

The use of flat space formulae for the black hole production
implies that the Schwarzschild radius~\eqref{214} is much less
than the AdS$_5$ curvature as viewed on the visible brane, $R_S
\ll \kappa^{-1}$. In its turn, this inequality means
\begin{equation}\label{216}
M_{bh} \ll 3\pi^2 \, x_1 \, \frac{\Lambda_{\pi}^2}{m_1},
\end{equation}
where $m_1$ is the mass of the lightest KK gravitons, and $x_1$ is
the first zero of the Bessel function $J_1(x)$. Let us stress, the
inequality for $M_{bh}$ in this form~\eqref{216} is valid in both
scheme \eqref{04} and scheme \eqref{16}. Since in our case with
$\kappa \ll \bar{M}_5 \sim 1$ TeV the value of $\Lambda_{\pi}$ is
of the order of $100$ TeV, while the lightest mass is about 60 MeV
(see estimates after Eq.~\eqref{24}), inequality~\eqref{216}
admits much higher $M_{bh}$ than one can have in a usually adopted
scheme with $\kappa \sim \bar{M}_5 \sim \bar{M}_{Pl}$, in which
$m_1 \sim \Lambda_{\pi} \sim 1$ TeV.

In Fig.~\ref{fig:sigma_in_bh} we present the black hole production
cross section in comparison with the gravitational cross section.
As one can see, at $\bar{M}_5 = 0.5$ TeV, gravireggeon
interactions (middle solid line in Fig.~\ref{fig:sigma_in_bh}) can
dominate black hole production mechanism at $E_{\nu} \gtrsim 4
\cdot 10^9$ GeV (dash lines).

\section{Conclusions and discussions}

In order to get a correct interpretation of the KK graviton masses
in the RS-like model, one can use the non-factorizable metric
which has an exponentially decreasing warp factor $\exp(-2\kappa
|y|)$ in its 4-dimensional part and then turn to the Galilean
coordinates. The SM fields are assumed to be placed on the TeV
(visible) brane located at $y = \pi r$, where $r$ is the size of
the 5-th dimension. Remember that $\kappa$ is a measure of the
negative constant curvature of the AdS$_5$ space.

Another way is to choose the exponentially growing warp factor,
namely $\exp(2\kappa |y|)$, but to place the visible brane at the
point $y=0$~\cite{Giudice:04}. In such a case, the coordinates are
Galilean from the very beginning. This choice of the warp factor
is equivalent to a formal replacement $\kappa \rightarrow
-\kappa$. Both ways lead to the hierarchy relation $\bar{M}_{Pl}^2
\simeq (\bar{M}_5^3/\kappa) \, \exp(2\pi \kappa r)$. Thus, one can
get TeV-phenomenology even if $\kappa \ll \bar{M}_5$, due to a
presence of the large factor $\exp(2\pi \kappa r)$, provided
$\bar{M}_5 \sim 1$ TeV and $\kappa r \approx 10$.

In the present paper, we have considered the case $\kappa \ll
\bar{M}_5 \sim 1$ TeV and have studied the inelastic scattering of
the brane fields induced by gravitational interactions in
$t$-channel. Namely, we have summed an infinite set of
trajectories (gravireggeons) corresponding to the massive KK
gravitons which lie on these trajectories. The imaginary part of
the eikonal, $\text{Im}\,\chi(s,b)$, has been analytically
calculated. It coincides with the imaginary part of the eikonal
derived in the scheme with one flat extra dimension of the size
$R_c$~\cite{Kisselev:04}, after a replacement $R_c \rightarrow
(\pi \kappa)^{-1}$. It is interesting to note that
$\text{Im}\,\chi(s,b)$ depends on the 5-dimensional Planck scale
$\bar{M}_5$ and the slope of the gravireggeons  $\alpha'_g$, but
it does not depend on $\kappa$ in the limit $\kappa \ll \bar{M}_5$
(up to negligible corrections).

Thus, the scattering of the SM particles in the AdS$_5$ space with
a small curvature looks similar to their scattering in the
5-dimensional flat space. It does not mean, however, that the RS
model with small curvature is equivalent to the ADD model with
\emph{one} large extra dimension of the size $R_c^{-1} = (\pi
\kappa)$. Indeed, in the ADD model with the fundamental scale of
order of one TeV, $R_c^{-1} \sim 10^{-30/d + 6}$, where $d$ is the
number of compact dimensions. According to this relation,
compactification radius $R_c^{-1} = (\pi \kappa) \approx 50$ MeV
can be realized only for $d=7$.

The results has been applied to the calculation of the gravity
contribution to the scattering of ultra-high energy neutrino off
the nucleon as a function of  the neutrino energy $E_{\nu}$. In
particular, we have found that for $M_5 \simeq 1$ TeV (which is
equivalent to $\bar{M}_5 \simeq 0.5$ TeV) the gravitational part
of the inelastic cross sections appeared to be comparable with (or
larger than) the black hole production cross section for
$M_{bh}^{min} = 1 \div 2$ TeV in the region $E_{\nu} \gtrsim 4
\cdot 10^9$ GeV.

Note, in the model with the flat metric, gravireggeon cross
section for the neutrino-nucleon scattering grows significantly
for small $d$~\cite{Kisselev:04}. Unfortunately, small values of
the number of the flat dimensions ($d \leq 3$) are ruled out by
the astrophysical bounds~\cite{Hannestad:03}. On the contrary, the
scheme with the warped metric and one extra dimension is free of
these bounds, and rather large cross sections (up to $0.01-0.1$
mb, at $E_{\nu} = 10^{12}$ GeV) are expected in this case. The
neutrino-nucleon cross sections will be probed by the Pierre Auger
Observatory at the level of SM predictions, taking into account
the high statistics to be collected by this experiment in six
years of operations~\cite{Anchordoqui:04}.

%%%%%%%%%%%%%%
% References %
%%%%%%%%%%%%%%

\end{document}